\documentstyle[prl,aps,preprint,floats]{revtex}

\def\Journal#1#2#3#4{{#1}{\bf #2}, #3 (#4)}

\def\NCA{Nuovo Cimento}

\def\NPB{{Nucl. Phys.} B}
\def\PLB{{Phys. Lett.}  B}
\def\PRL{Phys. Rev. Lett.}
\def\PRD{{Phys. Rev.} D}
\def\ZPC{{Zeits.f. Phys.} C}
\def\ZPA{{Zeits.f. Phys.} A}
\def\JPG{{Jour. Phys.} G}

\def\II{\relax{\rm I\kern-.18em I}}
\def\IH{\relax{\rm I\kern-.18em H}}

\def\frac#1#2{{#1\over#2}}

\begin{document}


\tighten

\title{FRAGMENTATION FUNCTIONS FOR BARYONS IN A QUARK-DIQUARK MODEL\thanks
{This work is supported in part by funds provided by the U.S.
Department of Energy (D.O.E.) \#DE-FG02-92ER40702.}}

\author{Anatoly D. Adamov and Gary R. Goldstein}

\address{Department of Physics \\
Tufts University \\
Medford, MA 02155 \\
{~}}

\date{TUFTS TH-97-G01 \hspace{0.3in}   Submitted to: {\it Phys. Rev. D}
  \hspace{0.3in} June 1997}

\maketitle

\begin{abstract}
A perturbative QCD calculation of heavy flavor quark fragmentation into 
heavy flavor baryons is developed along the lines of corresponding 
heavy meson models. The non-perturbative formation of the baryon  
is accomplished by implementing the quark-diquark model of the baryons. 
Diquark color form factors are used to enable the integration over the
virtual heavy quark momentum.
The resulting spin independent functions for charmed and bottom quarks to
fragment into charmed and bottom baryons with 
spin 1/2 and 3/2 are compared with recent data. 
Predictions are made for the spin dependent fragmentation functions as 
well, particularly for the functions $\hat{g}_1$ and $\hat{h}_1$ in the 
case of spin 1/2 baryons.

\end{abstract}


\section{Introduction}

Quarks produced in high energy processes materialize by evolving into
jets of hadrons. The particular hadronic fragments and their kinematic
dependences are of considerable interest. This fragmentation process
reveals the features of the non-perturbative regime of QCD. The
inclusive process of quark fragmentation into a single observed hadron
along with any number of unobserved accompanying particles is described by
a set of fragmentation functions. These fragmentation functions, defined
in terms of appropriate light cone variables, kinematic variables and
invariants, are probability distributions. These functions have received
considerable attention in
recent years. While experimental information beyond the pion distribution
(presumably from light quarks) has been slow in accumulating, theoretical
interest has been growing. The particular functional form for heavy
flavored quarks to fragment into heavy flavored hadrons is of special
interest. 

Experimentally it is possible to identify heavy flavored jets by
the production and characteristic flavor changing weak decays of the
hadrons. Theoretically this situation has been studied using
Operator Product Expansion techniques, light cone quantization, QCD
perturbation theory, and Heavy Quark Effective Theory, among other
methods.
Some of these methods yield general properties that reflect the overall
structure of QCD, as it is currently understood. Other approaches take
particular models of the low energy behavior expected from QCD, but in
regions that are not perturbatively calculable. The particulars of the
various approaches are near the point of being tested against
experiment. One general feature is known - 
the peak of the hadron distribution moves toward higher momenta as the 
quark mass increases. This feature is a result of the kinematics implicit 
in most models of the non-perturbative process, and is incorporated in the 
phenomenological Peterson function~\cite{peterson} that is used by
experimenters to fit 
the sparse data on heavy quark fragmentation~\cite{CLEO}.

It is more difficult to test the spin dependences of the
fragmentation processes experimentally. Yet these dependences are
very important to know. They
reflect the details of the primarily non-perturbative mechanism by which parton
polarization is passed on to the hadrons. The
spin-dependent fragmentation involves the reverse of the process by which
the nucleon spin is shared by its partons (the``spin crisis''), and may
reveal a similarly mysterious decoupling of valence quark spin and hadron
spin for some regions of kinematics. 

For the fragmentation of a heavy flavor quark into ``doubly heavy''
mesons, perturbative QCD may provide a starting point to a
theoretical determination of the fragmentation functions. As examples,
the fragmentation of a c-quark into the $J/\Psi$ or the b-quark into the
$B_c$, were calculated several years ago~\cite{chang,braaten}. 
The reasoning follows the observation that a heavy quark or diquark pair
must be produced to form the final hadron. At least one member of the pair
will be nearly
on-shell because the heavy flavor hadrons have small binding energies
relative to their masses. So, the gluon producing the heavy pair must
carry large squared time-like 4-momentum, $k^2$. This gluon will be shaken
off by the virtual fragmenting quark and the relevent coupling will be of
order $\alpha_s(k^2)$, which will be small. Hence perturbative QCD will be
applicable. Non-perturbative effects will be incorporated into the binding
of the initial quark with the pair produced heavy quark or diquark. If
this is the case, the fragmentation functions are
calculable, at an appropriate scale. The parton shower that accompanies
the jet is a result of QCD radiative corrections, which can be
obtained from the Renormalization Group or the Altarelli-Parisi equations.

Such calculations have been performed and scrutinized. It has been shown
that in the heavy quark limit (i.e. the mass goes to infinity) the
functions have the form expected from more general 
considerations~\cite{randall}. This
corresponds to the heavy meson taking all of the heavy quark's momentum;
the distribution becomes a delta function at $z=1$. The $1/m_Q$ 
corrections are calculated also. In
any case, this approach can predict the spin-dependent fragmentation
functions along with their momentum and mass dependences. In the heavy 
mass limit, of course, the spin of the heavy quark is conserved, so the 
spin dependence is simple. What is of phenomenological interest is the 
next order correction, at least, since that has non-trivial spin 
dependence.

The spin dependences of fragmentation are most readily 
studied experimentally by observing baryons rather than mesons. This is 
true for the production of hyperons or heavy hyperons ($\Lambda_c$,  
$\Lambda_b$, etc.), wherein the weak, parity violating decays provide  
polarization analyses~\cite{chen}. 
To consider fragmentation into baryons in this perturbative scheme, the 
three quark system has to be confronted. A simple alternative is to 
consider the baryons as quark-diquark bound states~\cite{gold1}, and to
use the same 
perturbative method as for the mesons. In order for the perturbative 
calculation to be useful the creation of a heavy pair of quarks or 
diquarks must be an intermediate step. Ideally then, doubly heavy
baryon fragmentation would be an appropriate testing ground for this
scheme. Such data is sparse, however.

To begin to see the structure it will be 
worthwhile to stretch the region of applicability to the ``singly'' heavy 
baryons. We have been carrying out this program to see the expected spin 
and kinematic dependences, with the hope of providing an experimentally 
testable model~\cite{gold2}. A similar approach has been developed
independently by Martynenko and Saleev~\cite{russians}, but with
significantly different assumptions about how to represent the diquark
structure. Of immediate interest is the question of whether the 
baryon fragmentation functions have the same kinematic dependence as the 
meson case. In general the answer is no in this model. Secondly, does this
approach give the right magnitude for fragmentation into heavy
flavor baryons? With our careful specification of the diquark
chromodynamic form factors (unlike Martynenko and Saleev) the answer is
yes. The spin dependent fragmentation is 
interestingly distinct from the naive heavy quark limit in detail. The
calculations and results will be presented below, along with a 
comparison with some recent data.

\section{Perturbative calculation}

The first calculations of the fragmentation functions in the perturbative 
scheme were applied to some of the inclusive heavy flavor meson decays of 
the $Z^0$, as produced 
at LEP~\cite{chang,braaten}. The partial width for the inclusive decay
process
$Z^0\rightarrow H + X$ can be written
in general for any hadron H as
\begin{equation}
d\Gamma(Z^0 \rightarrow H(E) + X) = \sum_i\!\int_0^1\!dz\,d\hat{\Gamma}\!
(Z^0 \rightarrow i(E/z) + X,\mu)\, D_{i\rightarrow H}\!(z,\mu),
\label{eq:general}
\end{equation}
where H is the hadron of energy E and longitudinal momentum fraction z 
relative to the parton {\it i}, while $\mu$ is the arbitrary scale whose 
value will be chosen to avoid large logarithms. The quark and the hadron
can carry spin labels as well, and appropriate spin-dependent
fragmentation functions will be included, as we will show later.
The fragmentation function $D_{i\rightarrow H}\!(z,\mu)$ enters here in a
factorized form (that can be maintained through the evolution equations).
Upon obtaining the $D_{i\rightarrow H}\!(z,\mu)$ in the model to be
described, its evolution to observable scales is developed through summing
leading logarithms via evolution equations.

\begin{figure}
\vspace{8.0in}
\includegraphics{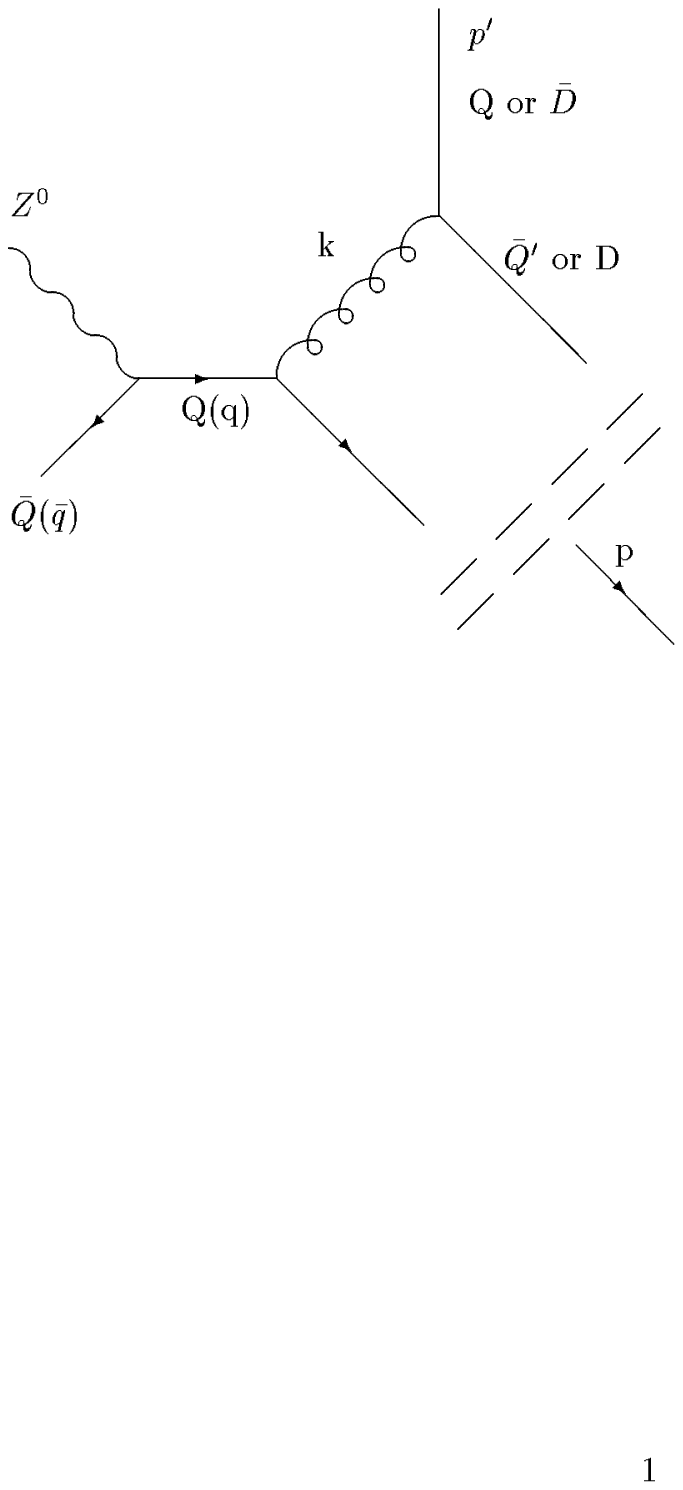}
\caption{The amplitude for $Z^0 \rightarrow$
Meson$(Q\bar{Q}')+X$ or Baryon$(QD)+X$.}
\label{fig:fig1}
\end{figure}

Now, consider the final state with one heavy flavor meson, say the $B_c$ 
for definiteness. We will soon replace this process by one involving a
heavy flavor baryon.
To leading order the $B_c$ meson arises from the production of a pair of 
b-quarks, in which one of the quarks fragments into the meson. As
Fig.~\ref{fig:fig1} 
illustrates, with $Q=b, Q'=c, \bar{Q}'=\bar{c}$, the perturbative
contribution involves the virtual b-quark
radiating a hard gluon (We work in axial gauge, so that there is no
contribution from the opposite quark). The hard gluon produces a heavy 
flavor pair of 
c-quarks. The gluon must have energy at least twice the charm mass, since
the $c$ and $\bar{c}$ are both on or near their mass shell. So 
the coupling $\alpha_s(k^2)$ is small, justifying the perturbative
approach. 

For nearly matching 4-velocities ($v=p/m$)
the $b$ and $\bar{c}$ form the $B_c$ meson bound state at roughly the
same 4-velocity, with amplitude
given by various projection operators multiplying the 
Bethe-Salpeter wave function $\chi(p,q_r)$ ($q_r$ is the $p(b)-p(\bar{c})$
relative momentum while p is their sum). Since the doubly heavy mesons  
are weakly bound objects (the sum of constituents' masses are near the
bound state mass), the wave function is expected to dampen non-zero
relative 3-momenta $\bf{q_r}$ in the hadron rest frame~\cite{guberina},
so that the $b$ and $\bar{c}$ 4-velocities are fixed at $p/M$. Then the
integration over the relative momentum of the two heavy quarks in
the full decay probability can be replaced by the squared amplitude
evaluated at equal 4-velocities (for the two constituents and the hadron).
The remaining integration over $\bf{q_r}$ applies to $|\chi(p,q_r)|^2$
which yields the square of the wave function at the origin in the hadron
rest frame. That wave function is known from non-relativistic 
quark models for the heavy-heavy meson system, or, more directly, from the
meson-to-vacuum decay constant. 

The same procedure can be applied directly to the baryons, if the 
quark-diquark model of the baryons is used. The hard gluon in the process 
must produce a diquark--anti-diquark pair, $D-\bar{D}$, and the diquark
(color anti-triplet) combines with the heavy flavor quark Q to form the
baryon $B_Q$. The relevent wave function will be
calculated from a fairly successful quark-diquark model~\cite{gold1}.

Note that an alternative scenario has been proposed by Falk, \textit{et 
al.}, in which the heavy quark fragments into a heavy diquark first, 
and then the diquark dresses itself to form the baryon~\cite{falk} with
probability of one for the latter. This 
leads to very different results, as pointed out in Ref.~\cite{russians}.
This scenario will not be used here, since the processes we are studying 
involve diquarks that do not necessarily carry the heavy flavor of
the quark. The latter authors~\cite{russians,russians2} have performed 
calculations that are similar in spirit to part of the procedure we follow
below, although not emphasizing the spin dependent structure functions
that we calculate below.

The tree level amplitude for Fig.~\ref{fig:fig1}, $A_1$, can be evaluated
explicitly
from perturbation theory.
The decay rate for unpolarized $Z^0 \rightarrow B_c+\bar{c}+b$ or
$B_Q+\bar{Q}+D$, each an exclusive channel, can be written generically as
\begin{equation}
\Gamma_1\,=\,\frac{1}{2M_Z}\!\int\![d\bar{q}][dp][dp']\,(2\pi)^4
\delta^4(Z-\bar{q}-p-p')\frac{1}{3}\!\sum\!|A_1|^2,
\label{eq:gam1}
\end{equation}
where $\bar{q}$, $p$, and $p'$ are the 4-momenta of the $\bar{b}$,
$B_c$ and $c$ (or the $\bar{Q}$, $B_Q$ and $\bar{D}$), respectively, and
$|A_1|^2$ is summed and averaged over unobserved
spins and colors. We use the notation $[dp]=d^3p/(16\pi^3p_0)$ for the
invariant
phase space element. In spin dependent fragmentation the sum will only
cover the unobserved outgoing parton spin labels ($c$ or $\bar{D}$ in this
explicit case).  To isolate the fragmentation function, the production of
the fragmenting quark ($d\hat{\Gamma}$ of Eqn.~\ref{eq:general}) must be
factored out. The fictitious decay width for
the $Z^0 \rightarrow b+\bar{b}$ or $Q+\bar{Q}$, with the $b$ or $Q$-quark
on shell is
\begin{equation}
\Gamma_0\,=\,\frac{1}{2M_Z}\!\int\![d\bar{q}][dq]\,(2\pi)^4
\delta^4(Z-\bar{q}-q)\frac{1}{3}\!\sum\!|A_0|^2.
\label{eq:gam0}
\end{equation}
with $q$ the $b$ or $Q$-quark 4-momentum.

To obtain the full inclusive width the unobserved quark degrees of freedom 
must be integrated over. By introducing the variables $q$, the off-shell 
quark's 4-momentum, and $s=q^2$, the square of the virtual mass, the two 
body phase space for $p$ and $p'$ can be written as an integration over
$z=(p_0+p_L)/(q_0+q_L)$ and $s$.
Note that the transverse momentum of the hadron, $\mathbf{p}_T$, and the  
unobserved quark, $\mathbf{p^\prime}_T = -\mathbf{p}_T$ (relative to the
fragmenting quark momentum), are
fixed for each pair of $z$ and $s$ values via the relation
\begin{equation}
s = q^2 = (q_0+q_3)(q_0-q_3) = \frac{M^2+{\mathbf{p}_T}^2}{z} +
\frac{{m^{\prime}}^2+{\mathbf{p}_T}^2}{1-z},
\label{eq:pT}
\end{equation}
where $M$ and $m'$ are the masses of the hadron and the unobserved
quark or anti-diquark, respectively. 
Then the phase space integration in  Eq.~\ref{eq:gam1}
can be written for the on-shell hadron
and unobserved quark and anti-diquark production as follows:
\begin{eqnarray}
\lefteqn{\int\![d\bar{q}][dp][dp']\,(2\pi)^4\delta^4(Z-\bar{q}-p-p')}
\nonumber  \\
  & &  = \int\!\frac{ds}{2\pi}\!\int\![d\bar{q}][dq](2\pi)^4
\delta^4(Z-q-\bar{q})\int\![dp][dp']\,(2\pi)^4\delta^4(q-p-p') 
\nonumber \\
  & &  = \frac{1}{16{\pi}^2}\int\!ds\!\int\![d\bar{q}][dq](2\pi)^4
\delta^4(Z-q-\bar{q})\int_0^1\!dz\frac{p_0}{zq_0}.
\label{eq:gam2}
\end{eqnarray}
The variables $p_0$ and $p_3$ have been replaced by s and z, and the
integration over ${\mathbf{p}}_T$ has been performed via the delta
function that requires
\begin{equation}
{\mathbf{p_T}}^2 = z(1-z)[s-s_{th}],
\label{eq:pT2}
\end{equation}
where $s_{th}=\frac{M^2}{z}+\frac{{m^{\prime}}^2}{1-z}$ is the minimum
value that $s$ can assume. Note that the azimuthal integration in
${\mathbf{p}}_T$ has been performed assuming there is no such dependence
in the amplitude. This will be true for spin averaged probabilities and
for products of helicity amplitudes, but not for other orientations of
quark or hadron spin. If spin projection operators are used in trace
expressions, care must be taken to integrate out the azimuthal dependence
first. The integrations over $q$ and $\bar{q}$ will be
common to the direct production of an on-shell quark in $\Gamma_0$ and the
off-shell quark that fragments in $\Gamma_1$. Providing the production 
dependence ($|A_0|^2$) can be factored out of the full probability
($|A_1|^2$), this will allow the fragmentation process to be defined
irrespective of the production mechanism, obviously an essential feature
of any model. The factorization will be possible in the appropriate large
momentum limit.

To match the integrand to the fragmentation function of
Eq~\ref{eq:general} the integration will be performed over the variable
$s$, keeping $z$ fixed and letting $M_Z$ and $q_0
\rightarrow\infty$. The $s$ integration
ranges from $s_{th}$ to $(M_Z-m_Q)^2$. Since the gluon
propagator in the amplitude emphasizes low values of $k^2$, and the heavy
quark propagator favors $s$
not far from on-shell; the major contribution to the $s$ integration
appears at low $s$. Thence, the upper limit of the integration over $s$ 
can be taken to $\infty$ to facilitate the evaluation of the definite
integral.
In the large $M_Z$ or $q_0 \rightarrow \infty$ approximation the 
transverse momentum of the hadron is small relative to $p_0$ and $p_3$,
since Eq.~\ref{eq:pT2} shows the transverse momentum is independent of
$q_0$ at fixed $s$ and $z$. Thus it is
sensible to ignore the transverse momentum in the relation $p=zq$ (after
carefully evaluating $s$ dependent terms in the integrand). Once
the square of 
the amplitude $A_1$ is summed over spins and simplified by dropping 
non-leading contributions, the width for $Z^0\rightarrow \bar{Q}Q$ can be 
factored out of the expression Eq.~\ref{eq:general} via
\begin{equation}
D_{Q \rightarrow H}(z)=\frac 1{16\pi
^2}\lim_{q_0\rightarrow \infty }
\int_{s_{th}}^{\infty}\! ds\frac{\left| A_1\right| ^2}{\left| A_0\right|
^2}
\label{eq:ratio}
\end{equation}
leaving an integral over the fragmentation function, since the production 
probability for the relevent quark has been factored out. Then
\begin{equation}
D_{Q\rightarrow H}(z)=\frac{8\alpha_s^2|R(0)|^2}{27\pi m_Q}
\!\int_{s_{th}}^{\infty}\!ds\!F(z,s),
\label{eq:frag}
\end{equation}
where $R(0)$ is the Bethe-Salpeter wavefunction at the origin, and
$F(z,s)$ is the remaining integrand, which depends on $s=q^2$, $z$ and the 
quark masses, with the $q_0\rightarrow\infty$ having been implemented. 
So the partial width for $Z^0 \rightarrow H + X$ is given by an 
integral over the virtuality of the heavy quark and the phase space of the 
unobserved degrees of freedom. 

We now proceed with the calculation of
fragmentation functions for (singly) heavy flavor baryons.
The basic covariant coupling of diquarks to gluons was written long 
ago~\cite{gold1}. There is one coupling constant for the scalar diquark 
color octet vector current coupling to the gluon field---a color charge 
strength, along with a possible form factor $F_s$. The momentum space color 
octet current (which couples to the gluon field vector) is
\begin{equation}
J_{\mu}^{A(S)} = g_s F_s(k^2) (p+p')_{\mu}S^{\alpha 
\dagger}\lambda_{\alpha \beta}^{A}S^{\beta},
\label{eq:scalar}
\end{equation}
where $p$ and $p'$ are the scalar diquark 4-momenta and $k=p'-p$.
For the vector diquark there 
are three constants - color charge, anomalous chromomagnetic dipole 
moment $\kappa$,
and chromoelectric quadrupole moment $\lambda$, along with the 
corresponding form factors, $F_E,\,F_M,\,{\rm and}\,F_Q$. 
\begin{eqnarray}
J_{\mu}^{A(V)} & = & g_s(\lambda^A)_{\beta 
\alpha}\left\{F_E(k^2)[\epsilon^{\alpha}(p)
\cdot\epsilon^{\beta \dagger}(p')](p+p')_{\mu}\right.\\
   & & +(1+\kappa)F_M(k^2)[\epsilon_{\mu}^{\alpha}(p)p
\cdot\epsilon^{\beta \dagger}(p')+\epsilon_{\mu}^{\beta
\dagger}(p')p'\cdot
\epsilon^{\alpha}(p)]\\
   & &
 +\frac{\lambda}{m_D^2}F_Q(k^2)[\epsilon_{\rho}^{\alpha}(p)
\epsilon_{\nu}^{\beta \dagger}(p')+\frac{1}{2}g_{\rho \nu}
\epsilon^{\alpha}(p)\cdot\epsilon^{\beta 
\dagger}(p')]k^{\rho}k^{\nu}(p+p')_{\mu}\left.\right\},
\label{eq:vector}
\end{eqnarray}
where $A$ is the color octet index, $\alpha,\,\beta$, ..., are color 
anti-triplet indices, the $\epsilon$'s are polarization 4-vectors for the
diquarks.

In the perturbative diagrams 
involved here, the virtual heavy quark emits a time-like off-shell gluon, 
that,in turn, produces a diquark-antidiquark pair while attaining nearly
on-shell 
4-momentum. The diquark combines with the heavy quark to form a heavy 
flavor baryon, whose amplitude for formation is related to the 
Bethe-Salpeter wavefunction for the diquark-quark system. As in the meson 
production calculations, it 
is assumed that the constituents are heavy enough so that the binding is 
relatively weak, i.e. the quark and diquark are both on-shell and the 
binding energy is negligibly small. This is expected to be true for 
constituents with masses well above $\Lambda_{QCD}$, and even the 
light flavor diquarks almost satisfy this constraint. The basic 
perturbative amplitude is shown in Fig.~\ref{fig:fig1} with the Q$'$-quark
line replaced by an (anti-)diquark D line. 

It should be realized that the integration (over $s$, the square of the 
virtual heavy quark mass) involved in the calculation 
would diverge for point-like vector diquarks, since the gluon coupling to 
a pair, Eq.~\ref{eq:vector} carries momentum factors. 
The virtual mass in the integration, $\sqrt{s}$, is passed on 
to the gluon and, subsequently, to the gluon-diquark vertex. 
Hence it is essential to regulate 
the integrand by some means. This is best accomplished via the 
chromoelectromagnetic form factors 
for the gluon coupling to the diquark. The 
form factor approach makes physical sense - it is a result of the 
compositeness of 
the diquarks. And for consistency, once the vector has form factors, the 
scalar diquark must have one also. 

There is no direct information about the chromoelectromagnetic form factors. 
We may expect that the ordinary electromagnetic form 
factors will have the same functional form as their QCD counterparts---the 
source of both sets of form factors is the matrix element of a 
conserved vector current 
operator. In the relevent case here, though, the vector operator is the 
gluon field --- a color octet. Also, what is of concern here is the 
time-like region of the form 
factor. For diquarks, of course, there is not any direct empirical 
evidence about their electromagnetic form factors, but diquark-quark models 
of the nucleon  have constrained the parameterization of the form factors. 
For one thing, 
the dimensional counting rules lead to $1/|q|^4$ asymptotic behavior of 
nucleon form factors (at asymptotic momentum transfer the baryon is a 
three quark system). A quark-diquark nucleon must approach this asymptotic 
behavior also, for consisitency. For a \textit{point-like} scalar diquark 
bound to a quark, the 
asymptotic behavior will be $1/|q|^2$ from dimensional counting for the 
exclusive pair production. Hence the composite scalar diquark must have an 
effective coupling to the gluon, i.e. a form factor, that approaches 
asymptotia as $1/|q|^2$. Since the vector particle has a 
polarization 4-vector associated with it, an extra power of momentum 
arises in the asymptotic amplitude. It becomes necessary for the charge 
and magnetic form factors to have $1/|q|^4$ asymptotic dependence, and the 
quadrupole $1/|q|^6$ behavior.

Using a quark-diquark model of the nucleon, and the well measured 
electromagnetic form factors of the nucleon, Kroll, {\it et 
al.}~\cite{kroll}, have obtained \emph{electromagnetic} form factors for
the diquarks. Two vector form factors are given $1/|q|^4$ asymptotic
behavior (the third is set to zero) and the scalar behaves as 
$1/|q|^2$. For the nucleon form factor study~\cite{kroll} as well as a
recent study of higher twist contributions to the 
nucleon structure functions~\cite{anselmino}, the scalar diquark form
factor and vector diquark form factor are assumed to have simple pole 
and dipole forms, respectively, with pole positions $M_S$ and $M_V$ 
above 1 GeV,
\begin{eqnarray}
\lefteqn{F_S(k^2)=1/(1-k^2/M_S^2),\ F_E(k^2)=1/(1-k^2/M_V^2)^2,}
\nonumber \\
 & &   F_M(k^2)=(1+\kappa)F_E(k^2),\ F_Q(k^2) = 0.
\label{eq:form}
\end{eqnarray}
The pole position values are somewhat higher than the dipole position 
for the overall nucleon form factors - near 800 MeV
If we make the assumption that the color 
form factors have the same functional form as the electromagnetic form 
factors, we can proceed. However,the region of most relevence for the 
fragmentation functions is time-like $k^2$, below the $4m_N^2$ 
threshold. In the $s$ integration that 
will be performed here, the time-like $k^2$ region begins at
$4m_{Diquark}^2$ 
for the value $z=1/(1+m_D/m_B)$, and at higher values for other choices of 
$z$. This implies that the integration region either overlaps or comes
near 
to overlapping the pole positions. The pole singularities have to be 
tamed, and the final integration may be very dependent on the method used 
to moderate the singularities. Treating the poles as real resonance 
positions, including a small imaginary part, on the order 
of the nearby vector meson width, would be sensible physically.
However, since the color octet form factor would be 
dominated by color octet vector mesons, and the latter are not expected to 
be strongly bound or narrow resonances, pole positions with  
large widths may be preferred. This would hide our ignorance and provide
an interpolation between the 
space-like and time-like asymptotic regions. That is the ansatz we adopt.

The neccessity for diquark form factors has an important consequence
theoretically. In the light cone expansion, the baryon
production via three quarks would contribute to the leading twist
fragmentation functions. Dimensional counting requires the $1/|q|^4$
behavior to which we alluded above. But the diquarks depart from this
behavior except at asymptotia. Hence the diquark form factors produce
non-leading twist behavior for the fragmentation functions. Gluon
contributions are buried in those form factors.
In terms of the full set of such functions~\cite{jaffeji}, we have more
than just
$\hat{f}_1$, $\hat{g}_1$ and $\hat{h}_1$. There are non-leading twist
functions like $\hat{e}_1$ and $\hat{h}_2$ that will receive contributions
at next-to-leading twist. When we determine what we call
$\hat{f}_1$, $\hat{g}_1$ and $\hat{h}_1$, we actually have some
non-leading twist contributions that have not been disentangled.

The amplitudes for the baryon production can now be calculated. The spin 
1/2 ground state baryons are composed of a scalar diquark and a heavy 
quark in an s-state. There is only one coupling, and it involves the 
$F_S$. The amplitude is
\begin{equation}
A_{S\,1/2}=-\frac{\psi(0)}{\sqrt{2m_d}}F_S(k^2)\bar{U}_Bg_s
[k_{\lambda}-
2m_d v_{\lambda}]P^{\lambda},
\label{eq:As}
\end{equation}
where 
\begin{equation}
P^{\lambda}=
\bigtriangleup^{\lambda \nu}g_s\gamma_{\nu} 
\frac{m_Q(1+\mathbf{v})+\mathbf{k}}{(s-m^2_Q)}\Gamma.
\label{eq:Plambda}
\end{equation}

For the vector diquark baryons, there are two form factors (we 
take the quadrupole to be zero -- it falls as $1/|q|^6$ asymptotically). 
The chromomagnetic coupling involves a parameter $\kappa$, the ``anomalous 
chromomagnetic moment''. This is taken to be -1.10, as will be explained
below in Section IV on comparing with data.
The s-state baryons are spin 3/2 and 1/2, which we will refer to as 1/2$'$. 
The 1/2$'$ lies between the 3/2 and the ground state 1/2 baryon. The 
amplitude for vector diquarks to be produced, along with the heavy quark, 
contributes to both 3/2 and 1/2$'$ states. The amplitude is conveniently 
divided into a chromoelectric and chromomagnetic part, involving the two 
distinct form factors. The chromoelectric part contributing to the spin 
1/2$'$ baryon is
\begin{equation}
A_{E\,1/2}=-\frac{\psi(0)}{\sqrt{3m_d}}F_E(k^2)\bar{U}_B\gamma_5\gamma^{\mu} 
\frac{1+\mathbf{v}}{2}g_s\epsilon^{\dagger}_{\mu}[k_{\lambda}-
2m_d v_{\lambda}]P^{\lambda}.
\label{eq:E1/2}
\end{equation}
The chromomagnetic contribution to the spin 1/2$'$ baryon is
\begin{equation}
A_{M\,1/2}=\frac{\psi(0)}{\sqrt{3m_d}}F_E(k^2)(1+\kappa) 
\bar{U}_B\gamma_5\gamma^{\mu} 
\frac{1+\mathbf{v}}{2}g_s[g_{\mu\lambda}(\epsilon^{\dagger}v)m_d-
\epsilon^{\dagger}_{\lambda}k_{\mu}]P^{\lambda}.
\label{eq:M1/2}
\end{equation}
For the spin 3/2 baryon the corresponding amplitudes are
\begin{equation}
A_{E\,3/2}=-\frac{\psi(0)}{\sqrt{2m_d}}F_E(k^2)\bar{\Psi}^{\mu}_B
g_s\epsilon^{\dagger}_{\mu}[k_{\lambda}-
2m_d v_{\lambda}]P^{\lambda},
\label{eq:E3/2}
\end{equation}
and
\begin{equation}
A_{M\,3/2}=\frac{\psi(0)}{\sqrt{2m_d}}F_E(k^2)(1+\kappa)\bar{\Psi}^{\mu}_B
g_s[g_{\mu\lambda}(\epsilon^{\dagger}v)m_d-
\epsilon^{\dagger}_{\lambda}k_{\mu}]P^{\lambda}.
\label{eq:M3/2}
\end{equation}
Each amplitude should be multiplied by the color factor $4/3\sqrt{3}$.
In these amplitudes, $\psi(0)$ is the Bethe-Salpeter 
wavefunction at the 
origin (for the s-state Q-diquark system), $m_d$ is the appropriate 
diquark mass, $U_B$ is a spin 1/2 Dirac spinor for the baryon, 
$\epsilon^{\dagger}(p')$ is the polarization 4-vector for the unobserved
anti-diquark, $\Psi^{\mu}_B$ is the Rarita-Schwinger spinor for the spin
3/2  baryon, $v=p/M$ is the 4-velocity for the 
heavy baryon of mass M, $\Gamma$ is the production vertex for 
the heavy quark--antiquark pair, $k$ is the 4-momentum of the gluon and, 
with $n=(1,0,0,-1)$, the corresponding propagator in axial gauge is
\begin{equation}
\bigtriangleup^{\mu \nu}=\frac{1}{k^2}(g^{\mu \nu }-\frac{n^\mu k^\nu 
+k^\mu n^\nu }{(nk)}). \nonumber
\end{equation}

Considerable simplification of these amplitudes follows. Recall that the
formation of the baryon requires that the quark and diquark carry the same
4-velocity as the baryon, so that $k=rp+p'$, with $r=m_D/M$ and
$m_Q+m_D=M$ in the weak coupling approximation. 
This leads to many
simplifying relations among the kinematic variables. Of particular
importance is the relation $k^2=r(s-m_Q^2)$, which ties the gluon
propagator to the heavy quark propagator as the $s$ integration is
performed. The resulting simplified forms for each of the amplitudes
become
\begin{eqnarray}
A_{S\,1/2} &=& \frac{\psi(0)}{\sqrt{2m_d}}F_S(k^2)\bar{U}_B
2g_s^2\left( \frac{2M^2(1-r)+M\not k}
{(s-m_Q^2)^2}-\frac{(np)}{(nk)(s-m_Q^2)}\right)\Gamma, \\
A_{E\,1/2} &=& -\frac{\psi(0)}{\sqrt{3m_d}}F_E(k^2)\bar{U}_B\gamma_5
\frac{2g_s^2}{M(s-m_Q^2)^2}[({\epsilon}^{\dagger}p)+
M\not{\epsilon}^{\dagger}] \nonumber \\
   & & *[2M^2(1-r)-2\frac{(np)}{(nk)}(kp)+M\not{k}]\Gamma, \\
A_{M\,1/2} &=& \frac{\psi(0)}{\sqrt{3m_d}}F_E(k^2)(1+\kappa)
\bar{U}_B\gamma_5 \frac{g_s^2}{rM(s-m_Q^2)^2}
\{-2(kp)(p\epsilon ^{\dagger })(1-r) \nonumber \\
   & & +2\frac{(n\epsilon ^{\dagger })}{(nk)}(kp)^2-2r(p\epsilon ^{\dagger
})\frac{(np)}{(nk)}(kp)
+(3r-2)M(p\epsilon ^{\dagger })\not k
+2M\frac{(n\epsilon ^{\dagger })}{(nk)}(kp)\not k
\nonumber \\
   & & -2rM(p\epsilon ^{\dagger})
(kp)\frac{\not n}{(nk)}+\ 2r(kp)M\not \epsilon ^{\dagger }-(kp)\not \epsilon
^{\dagger}\not k\}\Gamma, \\
A_{E\,3/2} &=& -\frac{\psi(0)}{\sqrt{2m_d}}F_E(k^2)\bar{\Psi}^{\mu}_B
g_s{\epsilon}^{*}_{\mu}2g_s^2
\left( \frac{2M^2(1-r)+M\not k}
{(s-m_Q^2)^2}\right.
\nonumber \\
    & & \left. -\frac{(np)}{(nk)(s-m_Q^2)}\right) \Gamma, \\
A_{M\,3/2} &=&
\frac{\psi(0)}{\sqrt{2m_d}}F_E(k^2)(1+\kappa)\bar{\Psi}^{\mu}_B
g_s^2\frac1{r(s-m_Q^2)^2}\{-2r(p\epsilon ^{\dagger })(kp)\frac
{n_\mu}{(nk)}  \nonumber \\
 & & -2(1-r)k_\mu(p\epsilon^{\dagger }) 
+2(kp)\frac{(n\epsilon ^{\dagger })}{(nk)}k_\mu  
-k_\mu \not \epsilon
^{\dagger }\not k\}\Gamma. 
\end{eqnarray}

There are 3 cases 
to consider for each heavy quark flavor---3 final state baryons. For the
two states resulting from the
vector diquark, the electric and magnetic amplitudes must 
%
be added together. Then for each baryon, the 
amplitude is squared and a trace is taken to sum over spins (including 
spin projection operators for the spin dependent cases). The analog of 
Eq.~\ref{eq:gam1} is obtained for each baryon. By carefully organizing the 
terms in the integrand, the width for the inclusive production of the 
virtual heavy quark can be divided out to yield the analog of 
Eq.~\ref{eq:frag} for each baryon. Finally the integration over $s=q^2$ 
can be performed numerically---the form factors make it difficult to 
write an analytic expression for each case. The resulting $z$ dependent 
fragmentation functions are the ``boundary'' functions, obtained at a
scale $\mu^2$ at the threshold $(2m_D+m_Q)^2$, with the strong coupling
constant $\alpha(\mu=2m_D)$. To consider higher momentum
scales, the Altarelli-Parisi evolution equations are used. This strategy
essentially sums the leading log contribution of the parton shower that is
generated by the off-shell heavy quark. 

We have taken some particular cases to illustrate the results. For the 
c(su) or c(sd) baryons, the $\Xi_c$ states, the diquark is given a mass of
0.9 GeV/c$^2$ and the ratio of diquark to hadron mass is $r=0.33$. Form
factor parameterizations are discussed in Section 4. The 
resulting function, $\hat{f}_1(z,Q^2)$ is shown in Fig.~\ref{fig:fig2} for
the boundary 
value at $\mu=2m_D+m_Q$ and
for $Q_0=5.5$ GeV, the jet energy obtained at CESR. The three s-states
lead to 
different behavior and overall probability. Note that the 
1/2 ground state is produced roughly as frequently as the 1/2$'$ while
the latter is produced about twice as often as the 3/2 state.
The observed 1/2 ground states are produced from the decays of these
vector diquark states approximately as often as they are directly
produced. 

In Fig.~\ref{fig:fig3} the corresponding fragmentation functions for
charmed states with
non-strange diquarks are shown. These functions are evolved to 45 GeV,
the jet energy attained at LEP. Fig.~\ref{fig:fig4} shows the b-quark
states with
non-strange diquarks, also evolved to 45 GeV. It is clear that these
spin-averaged fragmentation functions peak at high z, even after
evolution. The vector diquark baryon states have noticable secondary peaks
at the low scale, which get diluted at higher scales.The secondary peaks
arise from the polynomials in s in the chromoelectric and chromomagnetic
contributions and depend on the amount of the latter, as fixed by the
parameter $1+\kappa$. The shapes
of the curves are significantly different from the Peterson function
shape~\cite{peterson} and eventually should be distinguishable
experimentally. The peak position
moves towards higher z as the heavy quark mass increases, as expected from
heavy quark QCD. 

\section{Spin dependent fragmentation functions}

In the preceding, the spin orientations of the virtual quark and heavy
baryon have been summed. In general, however, there is a spin dependence
to the fragmentation process. For a spin 1/2 baryon there are two
fragmentation functions that characterize the leading twist spin
dependence, $\hat{g}_1(z)$ and $\hat{h}_1(z)$. They correspond, at the
parton model level, to the transfer from the quark to the baryon of
longitudinal polarization (or helicity) and transversity~\cite{chen}. As a
consequence of the development leading to Eq.~\ref{eq:ratio}, the spin
dependent fragmentation function $\hat{g}_1$ can be written in the form
\begin{equation}
\hat{g}_1(z)=D_{Q(+) \rightarrow H(+)}(z)-D_{Q(+) \rightarrow H(-)}(z)=
\frac 1{16\pi^2}\lim_{q_0\rightarrow \infty }
\int_{s_{th}}^{\infty}\! ds\frac{\left| A_{++}\right| ^2-\left|
A_{+-}\right|^2}{\left| A_{0+}\right|^2},
\label{eq:g1}
\end{equation}
where $|A_{+\lambda}|^2$ is the probability for the helicity +1/2 quark
to produce a helicity $\lambda$ baryon, with a sum over the unobserved
diquark degrees of freedom implied. The $A_{0+}$ represents the
corresponding production of an on-shell helicity +1/2 heavy quark, at
large $q_0+q_3$. The
reader may ask, how can a heavy quark flip its helicity in hadronization?
For the scalar diquark combining with the heavy quark, 
this has to be an effect that would vanish in the heavy mass limit,
relative to the spin independent fragmentation, i.e. $\hat{g}_1(z)$ would
coincide with $\hat{f}_1(z)$. For the corresponding vector diquark case
this need not be true, since the diquark can carry negative helicity
leading to the opposite helicity for the baryon.

The analogous transversity function requires the superposed helicity
states $(|+\,\rangle\ \pm\,i|-\,\rangle\,)/\sqrt{2} = |y_{\pm}\,\rangle$. 
\begin{equation}
\hat{h}_1(z)=D_{Q(y_+) \rightarrow H(y_+)}(z)-D_{Q(y_+) \rightarrow
H(y_-)}(z)=
\frac 1{16\pi^2}\lim_{q_0\rightarrow \infty }
\int_{s_{th}}^{\infty}\! ds\frac{\left| A_{y_+,y_+}\right| ^2-\left|
A_{y_+,y_-}\right|^2}{\left| A_{0y_+}\right|^2},
\label{eq:h1}
\end{equation}
Now in the model we are considering the transversity can not flip, since
there is no chirality change in the matrix elements. We have the
transitions $\frac{1}{2}^{+} \rightarrow \frac{1}{2}^{+} + 0^{+}$ or
$\frac{1}{2}^{+} + 1^{+}$, analogous to a virtual quark decay into a
baryon and a positive parity boson. To get a non-trivial result there
needs to be an opposite parity bosonic state as well. 

In Figs.~\ref{fig:fig5} and \ref{fig:fig6} the longitudinal fragmentation 
function $\hat{g}_1(z,Q^2)$ is plotted for the two spin 1/2 states with
the same physical parameters as in the preceding figures. Note 
that $\hat{g}_1$ is very similar in shape to the spin averaged case for
the $\Lambda$ states, as expected when the diquark is a scalar. This shows
that the helicity flip contribution is relatively small for the values of
$r=m_D/M$ relevent for the bottom and charm baryons. Hence the
longitudinal polarization of the heavy quark is passed on to the baryon.
But we will see that this spin preservation is less than 100\%.
Furthermore, the production of excited baryons will dilute that spin
preservation. The spin 1/2 $\Sigma$ states have very different behavior,
showing the importance of non-zero helicity for the vector diquarks.

The spin dependent fragmentation functions for the spin 3/2 baryons are 
even richer in complexity. There are seven such functions at leading 
twist, many of which will be accessible from the decay distributions of 
these states into the 1/2 state plus a pion. While these fragmentation
functions have not been classified in the light-cone expansion formalism,
it is clear that the number of independent leading twist functions
coincides with the number of forward amplitudes for parity
conserving elastic scattering of spin 1/2 on spin 3/2. In general there
will be 2(2S+1)-1 such leading twist fragmentation functions for a quark
to fragment into a spin S hadron. In the heavy quark limit the helicity of
the quark will be preserved in the hadron, so we expect the analog of
$\hat{g}_1$ to be near the spin averaged function. 

\section{Comparison with data}

The analysis of charmed baryon production data has been extensive at CESR
and, more
recently, at LEP. Fragmentation data now exist from the CLEO
collaboration~\cite{CLEO} for 
some of the $\Xi_c$ states. The $\Lambda_c$ is studied at both LEP and
CESR. Bottom fragmentation into baryons is now being studied at LEP.
CLEO, in
particular, has determined spin independent fragmentation functions 
for the 
lowest mass spin 1/2 and 3/2 states $\Xi_c$. Polarization asymmetry has
been measured for $\Lambda$ and $\Lambda_b$ at LEP. Production rates for
$\Lambda_c$ from c-quarks have been determined. All of this data provide a
testing ground for the model being proposed here.  

The parameters that enter our calculations of fragmentation
functions (and their integrals over $z$) are the masses of the
constituents,
the poles in the chromodynamic form factors, the widths of those poles,
and the anomalous chromomagnetic moment of the vector diquark. For the
diquark masses we take $m(ud)=0.6$ GeV and $m(us)=0.9$ GeV. The baryon
masses are taken from the data, so the ratios, $r=m_D/M$ of diquark to
baryon masses are determined thereby for each baryon. It is assumed that
the difference between a baryon mass and the constituent heavy quark
mass plus the diquark mass is negligible, in order that the Bethe-Salpeter
wave functions need be evaluated only at the origin. The poles that enter
the form factors of Eq~\ref{eq:form} for the (ud) diquarks are taken from
the electromagnetic form factors~\cite{kroll}, $M_S(ud)=1.8$ GeV and
$M_V(ud)=1.2$ GeV. The
full width at half maximum for the scalar form factor is set at 0.88 GeV
and, for simplicity, the vector form factor is assumed to have the same
width. Recall that the width is introduced so that there are no
singularities in the physical region of $s$, the virtuality of the
fragmenting quark, or, correspondingly, $k^2$, the square of the gluon
4-momentum. It will transpire that the fragmentation probabilities will
depend critically on the pole positions and width, since the integration
region is dominated by the lowest values of $s$, where the heavy quark is
nearly on mass shell. In that region the poles are nearby. The pole and
width parameters are expected to be different for the (su) diquarks. We
take a cue from the electromagnetic form factors of the charged $\pi$'s 
and $K$'s, where the charge radii are roughly in a ratio of 1.3:1,
corresponding to a pole postion that increases by 1.3 for the strange
meson. This is qualitatively understandable by analogy. The $\rho$ vector
meson contributes to the $\pi$ charge form factor, while the $\phi$
vector meson contributes to the kaon electric charge form factor. The
latter has a mass 1.3 times that of the
$\rho$. So, to fix the (us) diquark chromodynamic form factors we choose
an overall scale factor of 1.4 for the pole positions and the width,
slightly bigger than the meson case.

The anomalous magnetic parameter $\kappa_{EM}$ was determined~\cite{kroll}
for the (ud) vector diquarks to be a positive number - a result of fitting
the composite nucleon form factors. For the chromomagnetic case, however,
a negative value is preferred for $\kappa$ from calculations of the mass
spectrum of excited baryons~\cite{gold1}. Furthermore, using a
diquark-quark model of the nucleon to calculate the electromagnetic
charge radii and polarizabilities preferred a negative value for
$\kappa_{EM}$. It is unclear what value to take for this parameter, given
the divergence of different methods. 

We will determine a value for
$\kappa$ by optimizing our model predictions compared to data for the
ratio of production probabilities, $R(\Sigma_b)=\Sigma_b/(\Sigma_b +
\Sigma_b^*)$. In
the model, $\Sigma_b$ and $\Sigma_b^*$ are $b$+vector$\{u,d\}$ diquark
states of spin 1/2 and 3/2. The difference in production probabilities or
$\hat{f}_1(z)$ for these two states depends sensitively on $\kappa$. Using
the
measured value from DELPHI~\cite{delphi} of $0.24\pm 0.12$ for the ratio,
we choose
$\kappa = -1.10$. This makes the overall chromomagnetic coupling small and
negative ($1+\kappa = -0.10$). The reason for this small value is that
the ratio $R(\Sigma_b)$ would be exactly 1/3 from spin counting if there
were no chromomagnetic term at all; the 1/3 is compatable with the data.

The simplest states to study, from our point of view, are the
$\frac{1}{2}^{+}$ ground states, since they involve the scalar diquark.
For these states the integral over $z$ of $\hat{f}_1(z)$ should correspond
to the total production probability for producing the state from the
corresponding heavy quark. However, there are contributions to the same
probabilities from the excited states that decay into these ground states.
Consider the $\Lambda_c$ fragmented from a c-quark, for which
OPAL~\cite{opal}
measures $5.6\pm2.6\%$ and CLEO~\cite{CLEO}
finds $9.5\pm 1.3\%$. These measurements include directly fragmenting
$\Lambda_c$'s along with any state that decays into this ground state. The
$\Sigma_c$'s (both spin 1/2 and 3/2 states) decay strongly into $\Lambda_c
+ \pi$, so contribute to the
rate. With the parameters chosen, we find 0.5\% for the directly
fragmented $\Lambda_c$ and 3.3\% when the $\Sigma_c$'s are included. This
is consistent with the LEP data. For the analogous b-quark system we
have fixed the $\Sigma_b + \Sigma_b^*$ rate to be 4.8\%, consistent with
experiment~\cite{delphi}, $4.8\pm1.6\%$. The total
$\Lambda_b$ is then 5.8\%, comparing nicely with the
measurement~\cite{aleph} of $7.6\pm4.2\%$. These and the following results
are summarized in the Table below.

It is significant to note that we have obtained these sizeable baryon
fragmentation rates, in contrast with the similar model of Martynenko and
Saleev~\cite{russians}. Saleev~\cite{russians2} obtains only 0.2\% for the
$\Lambda_b$. This indicates the
importance of our form factors in getting the correct normalizations. 

Having confidence in the overall normalization, we have reason to trust
the full fragmentation functions. These are shown in Fig.~\ref{fig:fig3}
and \ref{fig:fig4} for
the c and b states. The input, unevolved ``boundary data'' show a large  
peak at high $z$ and a secondary peak at medium $z$. That fairly severe
behavior is moderated considerably after evolving to the scale of CESR or
LEP. But even at the LEP scale, the functions are distinguishable from the
Peterson function, being peaked at higher $z$ and more skewed. As
sufficient data is gathered, it will be possible to see such a difference.

\begin{figure}
\vspace{7.5in}
\includegraphics{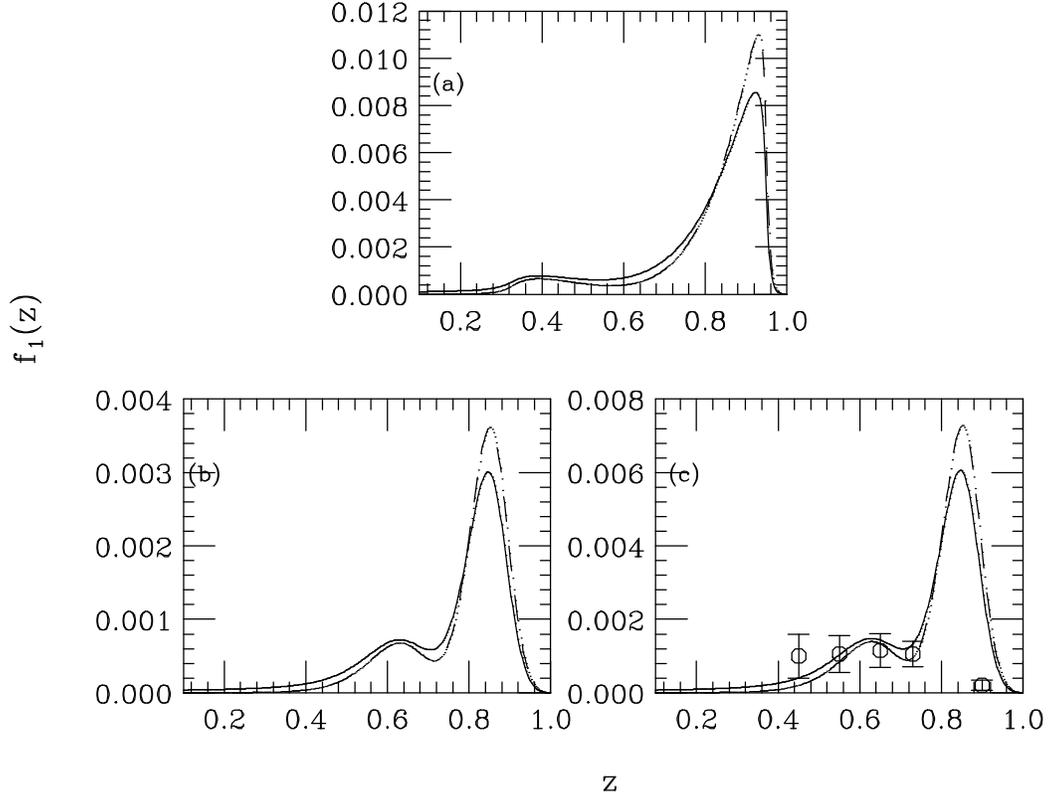}
\caption{Approximate $\hat{f}_1(z,Q^2)$ for a.$\Xi_c(1/2)$,
b.$\Xi_c^{'}(1/2)$, and c.$\Xi_c^{*}(3/2)$, each at $Q=\mu_0$ and $5.5$
GeV.}
\label{fig:fig2}
\end{figure}

\begin{figure}
\vspace{7.5in}
\includegraphics{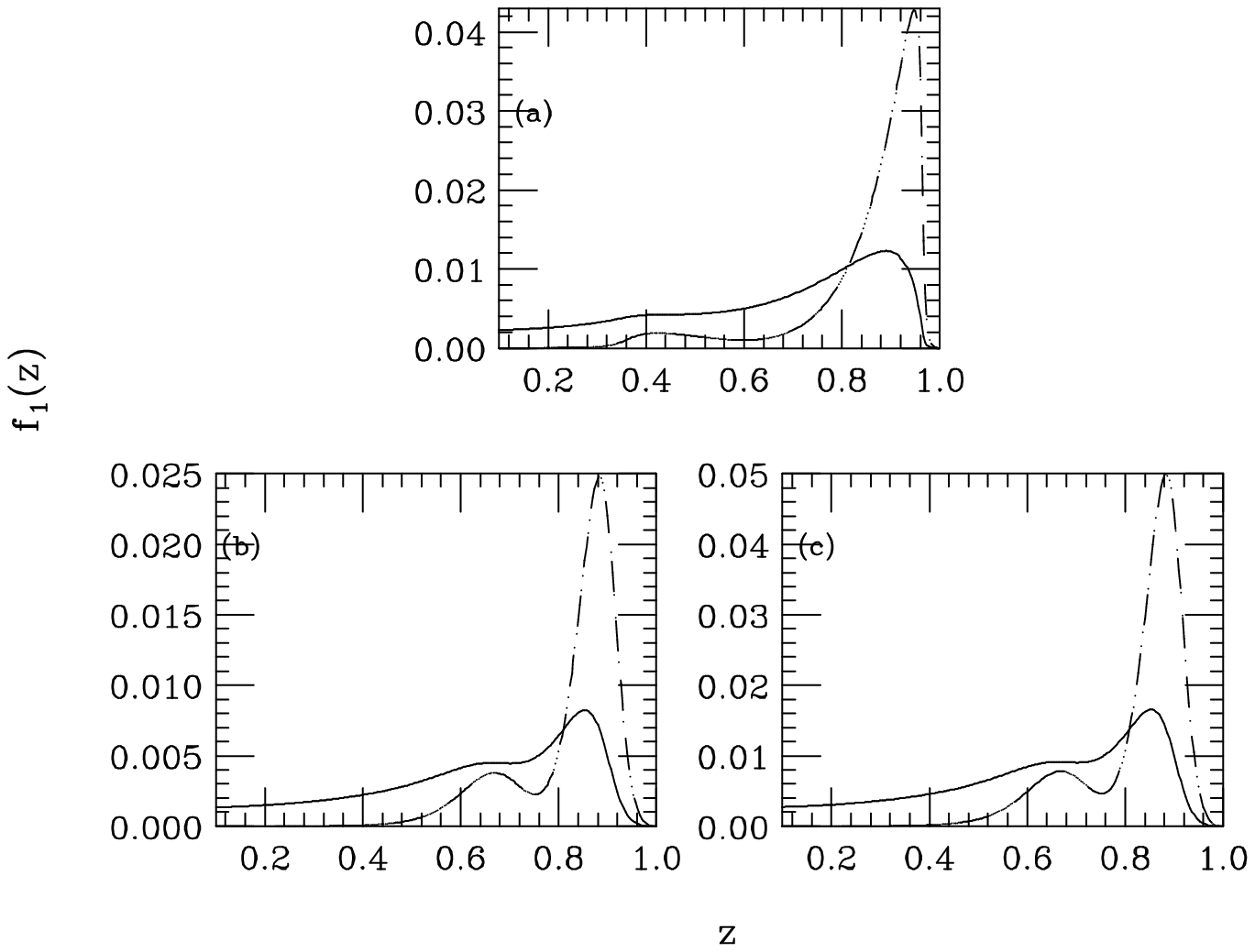}
\caption{Approximate $\hat{f}_1(z,Q^2)$ for a.$\Lambda_c$, b.$\Sigma_c$,
and $\Sigma_c^{*}$, each at $Q=\mu_0$ and $45$ GeV.}
\label{fig:fig3}
\end{figure}

\begin{figure}
\vspace{7.5in}
\includegraphics{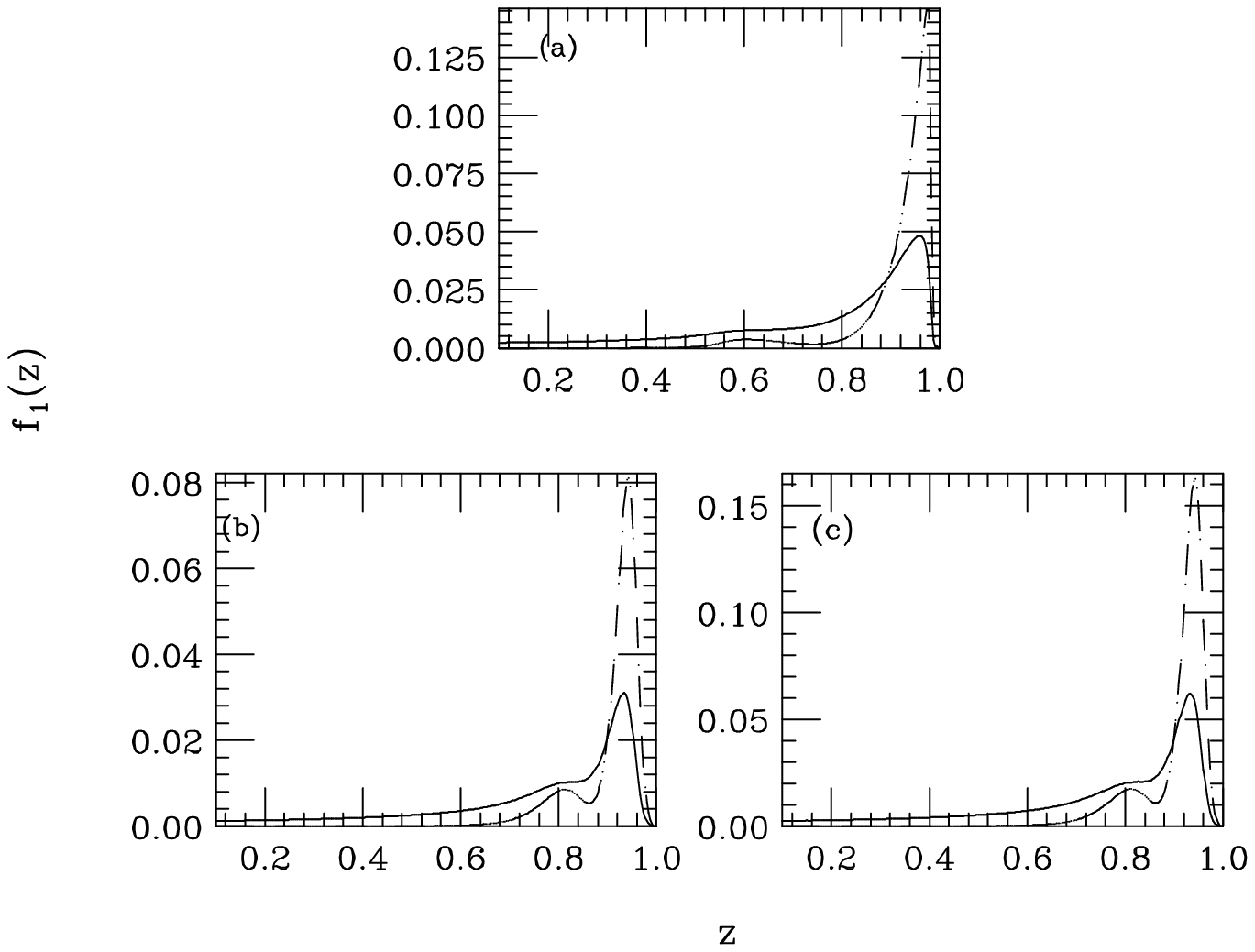}
\caption{Approximate $\hat{f}_1(z,Q^2)$ for a.$\Lambda_b$, b.$\Sigma_b$,
and $\Sigma_b^{*}$, each at $Q=\mu_0$ and $45$ GeV.}
\label{fig:fig4}
\end{figure}

\begin{figure}
\vspace{7.5in}
\includegraphics{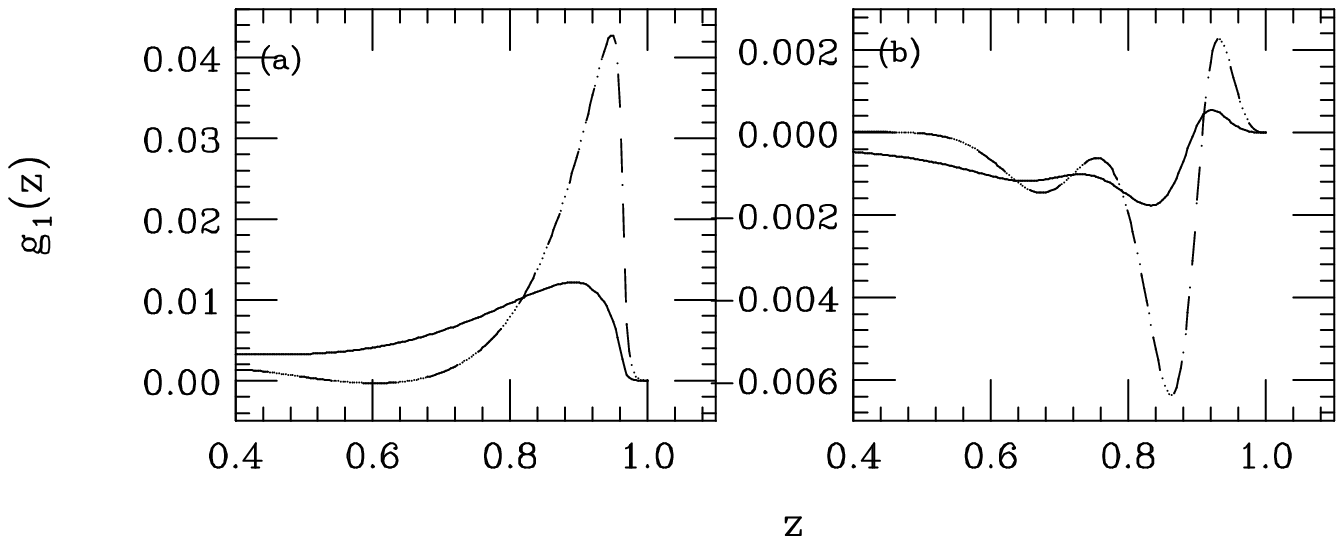}
\caption{Approximate $\hat{g}_1(z,Q^2)$ for a.$\Lambda_c$, b.$\Sigma_c$,
each at $Q=\mu_0$ and $45$ GeV.}
\label{fig:fig5}
\end{figure}

\begin{figure}
\vspace{7.5in}
\includegraphics{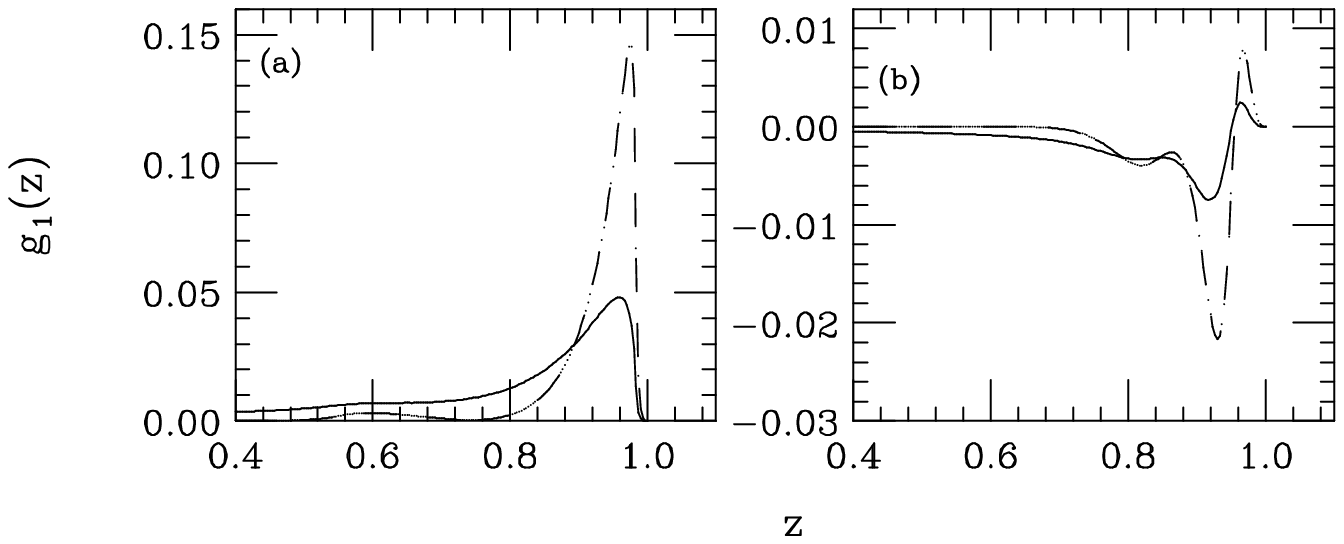}
\caption{Approximate $\hat{g}_1(z,Q^2)$ for a.$\Lambda_b$, b.$\Sigma_b$,
each at $Q=\mu_0$ and $45$ GeV.}
\label{fig:fig6}
\end{figure}

For the singly strange diquark, the lowest charmed baryon $1/2^+$ states
are the $c+[u,s]$ and $c+[d,s]$ states, $\Xi_c^{+}$ and $\Xi_c^{0}$, 
involving the antisymmetric, spin 0 diquarks. These, along with the  
spin $3/2^+$ states ($c+\{u,s\}$ and $c+\{d,s\}$ baryons), 
$\Xi_c^{*+}$ and $\Xi_c^{*0}$, involving the 
symmetric, spin 1 diquarks, have been seen and measured in sufficient
quantities for CLEO to sketch their fragmentation functions~\cite{CLEO}.
The spin $1/2^+$ partners, $\Xi_c^{'}$, of $3/2^+$ states 
have not been seen yet. They are presumed to have a mass below the 
$\Xi_c+\pi$ threshold, so must be seen in radiative decay channels. Note 
that these latter $\Xi_c^{'}\ 1/2^+$ states have the same isospin 
as the lower lying 
ground states $\Xi_c \ 1/2^+$ and could mix with them, in principle. In 
any case, the measured fragmentation functions provide a crude test of the 
model. The data are fit by the experimenters with a common
parameterization of the Peterson 
function~\cite{peterson}. It is easy to see in Fig.~\ref{fig:fig2} that
the data fall 
nicely on $\hat{f}_1$ of our model, evolved
to $Q=5.5$ GeV, with the possible exception of the highest z data point. 
These data are not 
sufficiently accurate to be a crucial test of the model, but do exhibit
the trends we expect. Note that the
experimental variable $x_p$~\cite{CLEO} does not correspond exactly to our
$z$, the light cone variable.

The ratio of the $3/2$ to $1/2$ production can 
be extracted from 
the data with some uncertainty~\cite{yelton}. The percentage of all 
$\Xi_c^{+}$ states that arose from decays $\Xi^{*0}_c \rightarrow 
\Xi_c^{+} + \pi^{-}$ is given as $(27 \pm 8)\%$ and the 
percentage of all 
$\Xi_c^{0}$ states that arose from decays $\Xi^{*+}_c \rightarrow 
\Xi_c^{0} + \pi^{+}$ is given as $(17 \pm 6)\%$. (Note that we have combined 
the statisitical and systematic errors here.) 

The experimenters do not see the $\pi^{0}$ channels, $\Xi^{*+}_c 
\rightarrow \Xi_c^{+} + \pi^{0}$ and $\Xi^{*0}_c 
\rightarrow \Xi_c^{0} + \pi^{0}$. From isospin conservation these channels 
account for 1/3 of the decays into $\Xi_c + \pi$, while the reported 
charged $\pi$ channels constitute 2/3. Suppose $N \, \Xi^{*}_c$ states of 
both charges are produced. Then 2/3 N will be seen in the charged $\pi$ 
decay mode. The total number of $\Xi_c^{+,0}$'s seen will be $N_{+,0} = 
\frac{2}{3}\!N/(0.27,0.17)$ (supressing errors until the end). The number 
of $\Xi_c^{+,0}$'s not coming from the decays of the 3/2 states will be 
$N_{+,0} - N$. Assume that $n_{+,0}$ of the $\Xi_c^{+,0}$'s come from 
other 
fragmented states' decays. Then $N_{+,0} - N - n_{+,0}$ is the number of direct 
fragmentation products of the charmed quark. The ratio $R(+\,{\rm or}\, 
0)$ of directly 
fragmented $\Xi_c^{+,0}$ to $\Xi^{*+,0}_c$ is given thereby as $R(+) = 1.5
\pm 
0.7 - n_{+}/N: 1$ and $R(0) = 2.9 \pm 1.4 - n_{0}/N : 1$. 

The numbers $n_{+,0}$ will come from the radiative decays of the heavier 
1/2 states, as well as higher $\Xi_c$ states (radial and orbital 
excitations of the $c+(su)$ and $c+(sd)$ systems). We have calculated the 
fragmentation functions for the spin 1/2 quark--vector-diquark states 
and hence the number of $\Xi_c^{'}$ spin 1/2$'$ states vs. $\Xi^{*}_c$
spin 3/2 states. That is 0.5:1 for the parameterization used in
Fig.~\ref{fig:fig2}. 
Assuming $n_{+,0}$ is due entirely to these $1/2'$ states decaying 100\% 
into the ground state $\Xi_c^{+,0}$, we have for the 
different charge states $R(+) = 0.9 \pm 0.7$ and $R(0) = 2.3 \pm 1.4$, 
both of which are consistent with the ratio of 1.4:1
predicted by the same model calculation. Hence, if the model is taken
seriously, and the experimental uncertainties are firm, the data do not 
require large contributions from fragmentation of the $c$-quark into
higher excitations of the $\Xi_c$ states. 

There are two reasons to be cautious about these experimental numbers,
however. First, the errors are quite large, leaving considerable variation
possible within two standard deviations. Secondly, the CLEO results are
obtained at $e^++e^-$ energy near 10 GeV. For the $c$-quark jet at roughly
5 GeV, the extraction of asymptotically meaningful fragmentation functions
is somewhat dubious.

In a previous version of our model~\cite{gold2}
our parameterization gave a much larger vector diquark to scalar diquark
production probability. With the more reasonable values now adopted these
diquark states and the corresponding baryons are produced with roughly the
same probabilities, as the 
calculations for heavy-heavy baryons by Martynenko and
Saleev~\cite{russians} favored.

Finally we consider the spin dependent fragmentation. At this time
there is not enough data to determine $z$ dependence for polarization in
heavy quark fragmentation. However, there is a determination of the net
longitudinal polarization of $\Lambda_b$ produced at LEP. That number is 
$-0.23\pm 0.25$ as determined by ALEPH~\cite{aleph}, using a technique
suggested by Bonvicini and Randall~\cite{bonvicini}. Given that the
b-quark produced at the $Z^0$ pole is expected to have longitudinal
polarization of $-0.94$, this measurement gives $0.24\pm 0.27$ for the
net transfer of helicity from the b-quark to the $\Lambda_b$. This is
rather low if one anticipates that the heavy baryon carries most of the
helicity of the heavy quark - the expectation of heavy quark field
theory~\cite{close}. 

Now the longitudinal polarization of the $\Lambda_b$ fragmenting from a
positive helicity b-quark is a function of $z$ - the ratio
$\hat{g}_1(z)/\hat{f}_1(z)$ for the directly produced $\Lambda_b$
(calculated to leading twist). The
integral over $z$ of that ratio would give a net polarization.
Experimentally, though, the net polarization is obtained by taking each
event, regardless of its $z$ (and $p_T$), and calculating a quantity
related to its polarization. The result of this process is to give a net
polarization that will be the integral of $\hat{g}_1(z)$ over $z$
divided by the corresponding integral of $\hat{f}_1(z)$. Both of these
integrals are scale independent. Using our fragmentation functions
we obtained 0.90, only marginally lower than the heavy quark
limiting value, but still not the small result extracted from the data.
However, from our spin
independent calculation above and the LEP data, we know that the
$\Sigma_b$ and $\Sigma_b^*$ are produced in relative abundance, and will
decay into $\Lambda_b+ \pi$, so that the polarization will be diluted by  
these other channels. A heavy quark limit calculation by Falk and
Peskin~\cite{peskin} anticipated this circumstance. They summed the
contributions of all of these states
in determining the net $\Lambda_b$ polarization. Using 
simplifying assumptions, they obtained about 0.72 for the fractional
helicity transfer from the b-quark to the hadron. That estimate assumed
the ratio (called A in their paper) of $\Sigma_b$ and $\Sigma_b^*$
(vector diquark states) to $\Lambda_b$ (scalar diquark states) of 0.45. 
Taking that ratio to be 4.8 instead, which is our result and consistent
with experiment, the resulting fractional helicity transfer becomes 0.26.
Falk and Peskin also define a parameter $w_1$ which measures the amount of
helicity $\pm1$ diquark that combines with the heavy quark. They take that
parameter to be zero, whereas we can calculate $w_1$, which averages over
all $z$ to be $\simeq 0.6$. With our values for both A and $w_1$ we obtain
the fractional helicity transfer of 0.46. Both of these results are
near the central value of the measurement. In obtaining these results we
have assumed the heavy quark limiting values of $\pm 1/3$ for the
polarization of the secondary $\Lambda_b$'s resulting from the $\Sigma_b$  
and $\Sigma_b^*$ decays. These results will obtain when the chromomagnetic
contribution is small, as it is for the $\kappa$ we have chosen.

The application of a similar model by Saleev~\cite{russians2} to
$\Lambda_b$ production and polarization yields very different results.
As we noted above, his production probability is too small. The
polarization for direct $\Lambda_b$ is less than ours (0.6 to 0.7).
However, it seems that it is the chirality asymmetry (Left - Right) that
Saleev has
calculated, rather than helicity. It is the latter that is measured
experimentally. We have a much weaker dependence on the diquark mass in
our calculation as a result.

\section{Summary}

A model for fragmentation into heavy flavored baryons has been developed
using perturbative QCD and a Bethe-Salpeter wave function for a
quark-diquark system. This provides the starting point for QCD evolved
fragmentation functions. The parameterization of the diquark structure
through the chromodynamic form factors for scalar and vector diquarks is
accomplished by using the pole form applied to the electrodynamic form
factors. The kinematic region near these poles is quite important because
the integral over the fragmenting quark's virtuality, $s$ (or indirectly,
the baryon's transverse momentum at a fixed $z$), emphasizes the 
nearly on-shell region where the corresponding gluon $k^2$ is near the
poles. It is this pole parameterization that is crucial for determining
the overall magnitudes of the various production probabilities. 

The production probabilities were all close to experimental values, which
supports our reasoning about the form factors. The $z$ and $Q^2$
dependences predicted have the common feature of being very sharply peaked
at high $z$ for the input scale $\mu_0$, and more broadly peaked at high
$Q^2$. For the 1/2$^+$ ground states, the $z$
dependence of the spin dependent $\hat{g}_1$ is close to the form for
$\hat{f}_1$, but their ratio (which will determine the baryon polarization
as a function of z) is striking. For the higher mass 1/2$^+$ state
containing the vector diquark the $\hat{g}_1$ has a very different $z$
dependence. It will be particularly interesting
to see if the peak and dip structures for both spin 1/2 states are
reproduced by the data. 

\section*{Acknowledgments}
This work was supported, in part by a grant from the US Department of 
Energy. G.R.G. appreciates the hospitality of the organizers of 
Diquark III where a preliminary version of this work was presented. We
appreciate helpful communications with J. Yelton regarding 
CLEO results and T. Behnke concerning OPAL results.



\begin{table}
\begin{tabular}{|l|l|l|}
Particle & Experiment & Prediction \\ \hline
$P(c\rightarrow \Lambda _c$) (including decays of $\Sigma _c$ and $\Sigma
_c^{*}$) & 5.6$\pm $2.6\% [OPAL] & 3.26\% \\ \hline
$P(b\rightarrow \Lambda _b)$ (including decays of $\Sigma _b$ and $\Sigma
_b^{*})$ & 7.6$\pm $4.2\% [ALEPH] & 5.8\% \\ \hline
$P(b\rightarrow \Sigma _b+\Sigma _b^{*})$ & 4.8$\pm $1.6\% [DELPHI] & 4.8\%
(fixed) \\ \hline
$\frac{P(b\rightarrow \Sigma _b)}{P(b\rightarrow \Sigma _b)+P(b\rightarrow
\Sigma _b^{*})}$ & 0.24$\pm $0.12 [DELPHI] & 0.33 \\ \hline
$\frac{P(c\rightarrow \Xi _c^{*})}{P(c\rightarrow \Xi _c)}$ & 2.3$\pm $1.4
and 0.9$\pm $0.7 [CLEO] & 1.4 \\ \hline
$P(c\rightarrow \Xi _c)$ (including decays of $\Xi _c^{\prime }$ and $\Xi
_c^{*}$) & - & 0.53\% \\ \hline
$P(c\rightarrow \Xi _c)$ (direct production) & - & 0.17\% \\ \hline
$P(c\rightarrow \Xi _c^{*+}$ or $\Xi _c^{*0})$ (direct production) & - & 
0.12\% \\ \hline
$P(c\rightarrow \Lambda _c$) (direct production) & - & 0.52\% \\ \hline
$P(c\rightarrow \Sigma _c^{*++}$ , $\Sigma _c^{*+}$ or $\Sigma _c^{*0})$
(direct production) & - & 0.62\% \\ \hline
$P(b\rightarrow \Lambda _b$) (direct production) & - & 1\% \\ \hline
$P(b\rightarrow \Sigma _b^{*+}$ , $\Sigma _b^{*0}$ or $\Sigma _b^{*-})$
(direct production) & - & 1.06\% \\ 
\end{tabular}
\end{table}

\begin{figure}
\vspace{7.5in}
\includegraphics{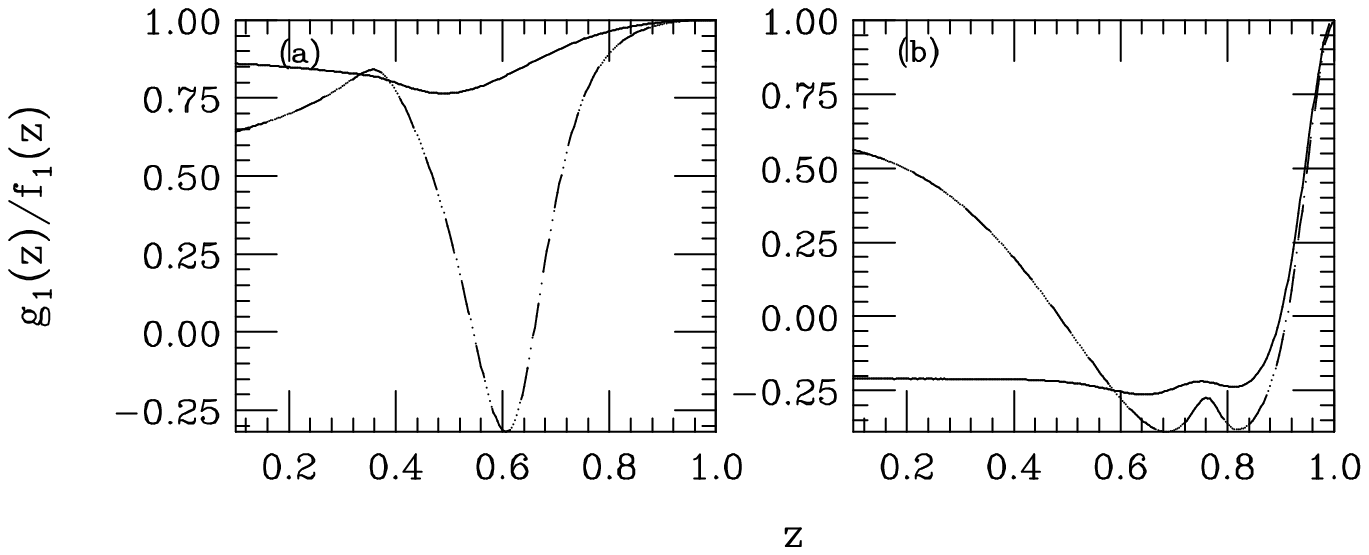}
\caption{Approximate ratio $\hat{g}_1(z,Q^2)/\hat{f}_1(z,Q^2)$ for
a.$\Lambda_c$, b.$\Sigma_c$,
each at $Q=\mu_0$ and $45$ GeV.}
\label{fig:fig7}
\end{figure}

\begin{figure}
\vspace{7.5in}
\includegraphics{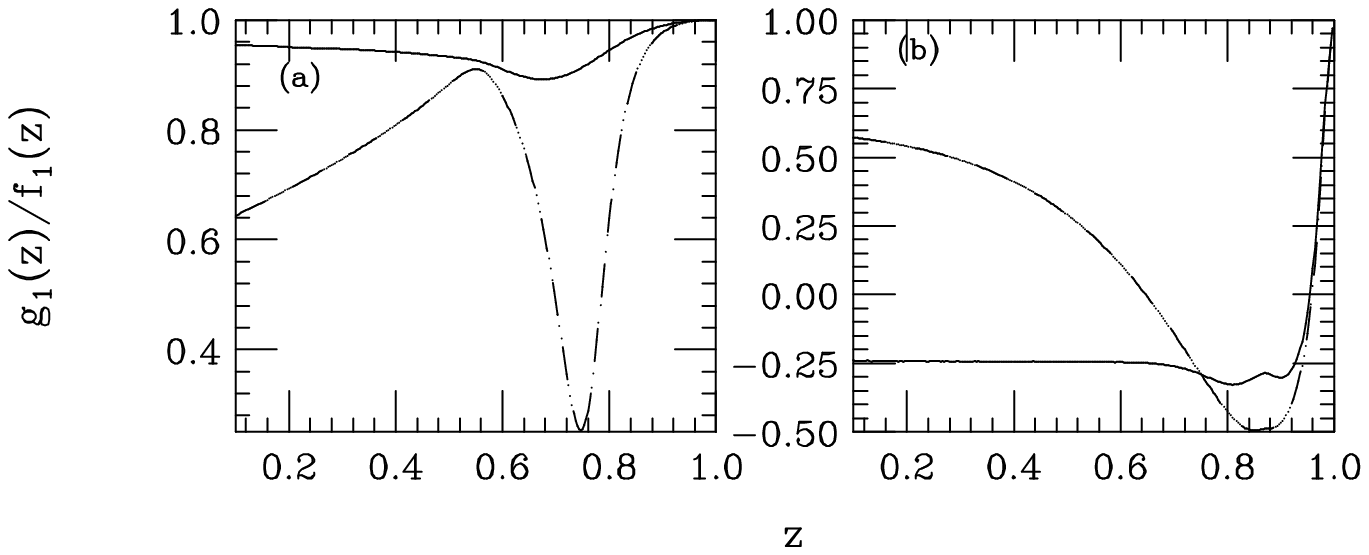}
\caption{Approximate ratio $\hat{g}_1(z,Q^2)/\hat{f}_1(z,Q^2)$ for
a.$\Lambda_b$, b.$\Sigma_b$,
each at $Q=\mu_0$ and $45$ GeV.}
\label{fig:fig8}
\end{figure}

\end{document}